# Defects induced polymer aggregates: A theoretical study

**Pramod Kumar Mishra**

Department of Physics, DSB Campus, Kumaun University, Nainital,Uttarakhand (INDIA)

**Abstract**. We consider three dimensional model of the Gaussian polymer chain in the presence of defects to understand the formation of a polymer aggregate where the aggregate is induced by the defects. The defects are acting as an attractive centres of the monomers and it induces aggregation of the monomers of the chain around the defects. It has been shown using the analytical calculations that the formation of a polymer aggregates are favoured when the defects have extensions in all the possible three dimensions. We have also calculated relevant other thermo-dynamical parameters (i. e. the average number of the monomers and the average size of the chain about the defect line) of the polymer aggregates to justify our findings.

## 1. Introduction

The Gaussian polymer model is widely used to understand the thermo-dynamical aspects of the macromolecules [1-2]. The macromolecules have complex structure and such complex structure forms the basis on interesting properties of the macromolecules. Thus, the properties of the macromolecules are related intimately to its structure. The Gaussian polymer model is a simple model to study analytically; therefore we have chosen the Gaussian model of a linear homo-polymer chain for the present study. The aggregate of the polymer chain may not have excluded volume interaction among its monomers and therefore, we have chosen the Gaussian polymer model to understand behaviour of the polymer aggregate in the presence of the defects.

The defects may be used to decorate the surface of a substrate; and a layer of the polymeric substance may be formed to protect the surface of the substrate from erosion [1-3]. The defects may be aligned along a line in one, two and three dimensions to form the polymer aggregate around the defects. We have chosen some of the possible geometrical arrangements of the defects and these defects are acting as the attractive centres for the monomers of the chain and a thick layer of the monomers (the nano polymer aggregates) may be formed around the defects. Since, the aggregation of the monomers of the chain may not have excluded volume interaction among its monomer and therefore the Gaussian polymer model may be suitable to understand thermodynamics of the polymer aggregate where the aggregate is induced due to geometrical arrangements of the defects.

The manuscript is organized as follows: we briefly describe the well known model of the Gaussian chain in the section-2 for chosen arrangements of the defects. The method of the calculations and the results are outlined in the section-3. The summary of our findings and the conclusions are illustrated in the section four.

## 2. The model

The Gaussian polymer model is known for its simple mathematics. It has been widely used to understand the thermodynamics of the polymeric systems [1-3]. Therefore, we have also used this model to understand the thermodynamics of the polymer chain in the vicinity of defects. The defects are assumed to aligned along a line, in a plane and also in three dimensional space; and their arrangement is such that we may form organization of a polymer aggregate; and the aggregate may have linear, planar or the volumetric extensions. Thus, we model the conformations of the chain as the Gaussian random walk and all directions ($\pm x$, $\pm y$ and $\pm z$) of the three dimensional space is permitted or in other words the walker is allowed to take steps in all the possible directions and there is no excluded volume interaction among the monomers of the chain [1-3]. The globular structure of the polymer chain (polymer aggregates) do not have excluded volume interaction [9]; and hence we use the Gaussian chain model to understand thermodynamics of the polymer aggregates. The aggregate of polymer chain is shown in the figure no. 1 schematically.

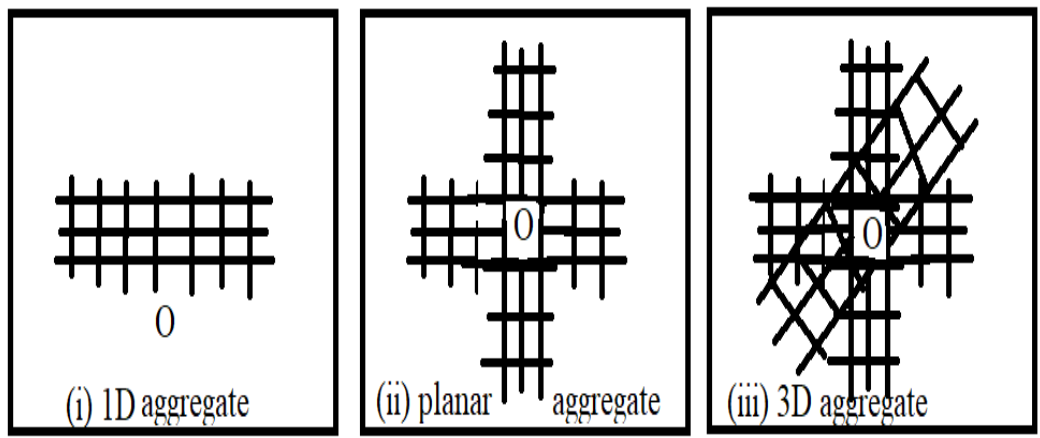

**Figure No. 1:** We have shown the schematic picture of an aggregate of the Gaussian polymer chain. The aggregate may have extension along a line 1 (i), in a plane 1 (ii) and also in the three dimensional space 1 (iii).

The grand canonical partition of the Gaussian chain for the proposed model system is written in general as,

$$Z(g,u) = \sum_{N=1}^{\infty} \sum_{\text{All walks of N monomers}} g^N u^m \qquad (1)$$

Where $g$ is the step fugacity of the monomers of the Gaussian chain in three dimensions and $u[=Exp(\beta E_s)]$ is the Boltzmann weight which corresponds to the onsite potential due to each defect. There are an $N$ monomers in the chain and a chosen conformation of the chain may has $m$ defects in it.

## 3. The method of calculations and the results

We write recursion relations [1, 8-15] to calculate partition function of the Gaussian chain where the monomers of the chain has attractive interaction with the onsite potential of the defects. Therefore, large number of the monomers are aggregated to form the nano polymer aggregate and the aggregate may have extension along a line, on the nano areal substrate or the aggregate may be in the three dimensions. We have written grand canonical partition function for each of the case chosen in the present report for the sake of discussion.

The singularity of the partition function is used to obtain the condition for the formation of the nano polymer aggregate, and we have shown the required value of the monomer-defects interaction energy for the formation of the aggregates in the figure no. 2. The figure shows that formation of aggregate require large interaction potential when defects are aligned along a line than the case where the defects are arranged in the three dimensional of two dimensional regions.

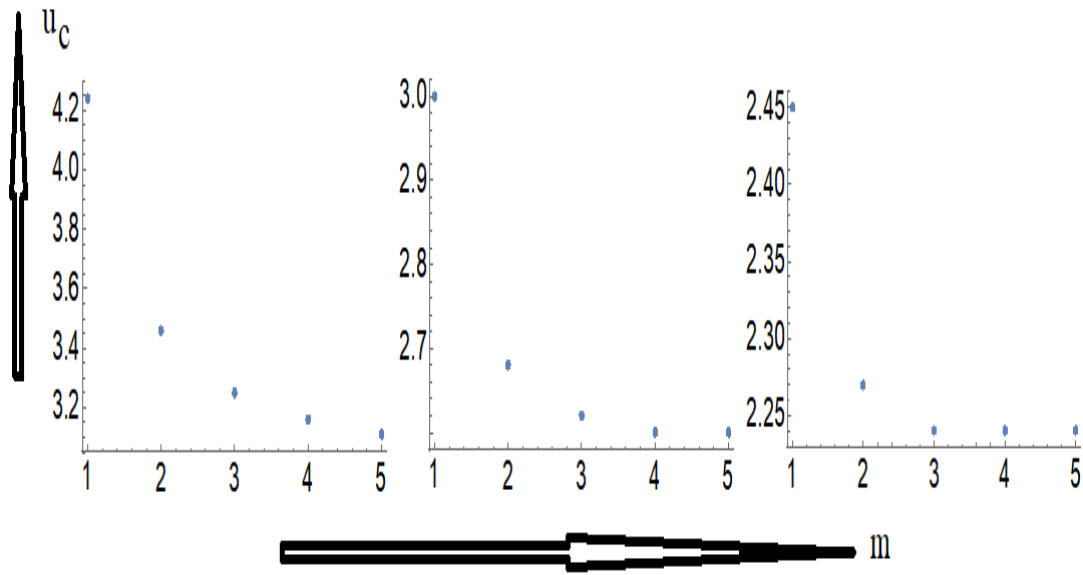

**Figure No. 2:** It is shown in this figure that we have polymer aggregate along a line, in a plane and also in the three dimensional space. The polymer aggregate is driven due to large entropy of the Gaussian chain and defects are helping to form the aggregate.

## 4. The summary and the conclusions

We used the Gaussian model [1-3] for the linear polymer chain in the presence of the defects, and the defects are acting as the attractive centres for the monomers of the chain. The defects are aligned along a line and these defects are arranged in one, two and three dimensions to form the nano polymer aggregates of different shapes. The organization of defects are schematically shown in the figure no. 1. A method of recursion relations is used to solve the proposed model [1, 8, 10-15] of defect induced polymer aggregates analytically to estimate the critical values of the monomer-defects interaction energy where the interaction energy is required to form the polymer aggregates of various shapes.

The results on the polymer aggregates is shown in the figure no. 2 regarding critical values of $U_c$ for one, two and three dimensional polymer aggregates. We have also calculated density of monomers along the defect line to understand geometrical shape of the aggregate.

We have shown through analytical calculations that there are 33.33% monomers of the chain may be located along an axis (i e. say along x-direction) and the average number of the monomers increases around defects as the attractive potential due to the defects are increased and thus, the average number of the monomers along the defect line grows to the maximum value as the defect potential is increased to $U_c$. The results on the defects induced monomers aggregation are shown in the figure nos. 3(i-vi).

The average number of the monomers along a line different from the defects line decreases from 33.33% to zero value as the defect potential is increased to $U_c$. Thus, there is aggregation of the monomers along the defect line. Our analytical estimates on the defects induced nano polymer aggregate formation show that the three dimensional aggregates are easily formed than two and one dimensional polymer aggregates.

The average number of the monomers along the defect line increases as the defect potential increased to $U_c$ value. The method of calculations proposed in the present report may be extended to study thermodynamics of a finite chain polymer aggregates.

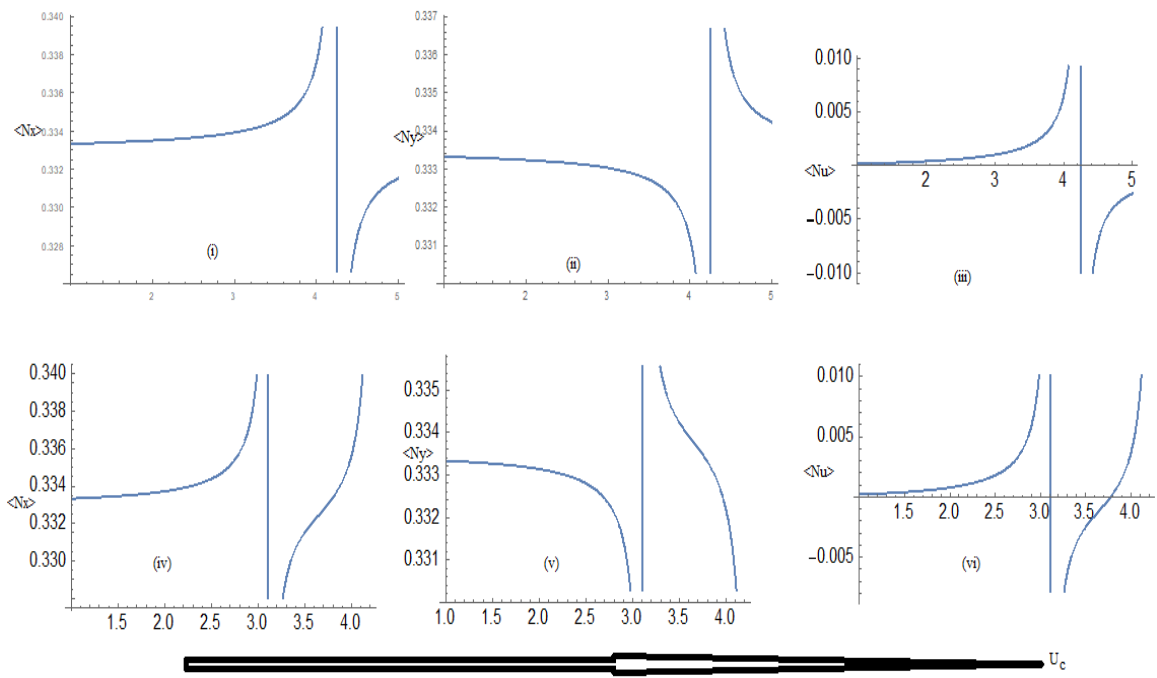

**Figure No. 3:** The monomer number density is shown in this figure to understand role of defect in the formation of polymer aggregates. It is seen from figure that number of monomer along defect line increases as the defect potential is increased. However, the number of monomer along other direction from defect line reduces as the potential is increased. One defect related results are shown in the figure 3(i-iii) and 3(iv-vi) corresponds to five defects along a line.